\documentclass[12pt,preprint]{aastex}
\newcommand{\beq}{\begin{equation}}
\newcommand{\eeq}{\end{equation}}

\newcommand{\cs}{c_{\rm s}}
\newcommand{\cmc}{~{\rm cm}^{-3}}
\newcommand{\cms}{~{\rm cm}^{-2}}
\newcommand{\Alf}{Alfv\'en\ }

\newcommand{\kms}{~\rm km~s^{-1}}

\newcommand{\pc}{~\rm pc}

\newcommand{\K}{~\rm K}
\newcommand{\muG}{~\mu{\rm G}}
\newcommand{\atil}{\tilde{a}_d}
\newcommand{\nutil}{\tilde{\nu}_0}
\newcommand{\lam}{\lambda_0}

\slugcomment{to appear in ApJ 2006 May}

\shorttitle{Hydromagnetic Wave Support}
\shortauthors{Kudoh \& Basu}

\begin{document}

\title{Nonlinear Hydromagnetic Wave Support of a 
Stratified Molecular Cloud \\ II: A Parameter Study}

\author{Takahiro Kudoh\altaffilmark{1} and Shantanu Basu}
\affil{Department of Physics and Astronomy, University of 
Western Ontario, London, ON N6A 3K7, Canada}
\email{kudoh@astro.uwo.ca, basu@astro.uwo.ca}

\altaffiltext{1}{Present address: National Astronomical Observatory of Japan,
Mitaka, Tokyo 181-8588, Japan; kudoh@th.nao.ac.jp}

\begin{abstract}
We use numerical simulations to study the effect of nonlinear 
MHD waves in a stratified, self-gravitating molecular cloud that 
is bounded by a hot and tenuous external medium. 
In a previous paper, we had shown the details of a standard model and
studied the effect of varying the dimensionless amplitude $\atil$ 
of sinusoidal driving.
In this paper, we present the results of varying two other important
free parameters: $\beta_0$, the initial ratio of gas to magnetic pressure
at the cloud midplane, and $\nutil$, the dimensionless frequency of
driving. Furthermore, we present the case of 
a temporally random driving force. 
Our results demonstrate that a very important consideration for
the actual level of turbulent support against gravity is the 
ratio of driving wavelength $\lam$ to the the size of the
initial non-turbulent cloud; maximum cloud expansion is achieved
when this ratio is close to unity.
All of our models yield the following basic results:
(1) the cloud is lifted up by the pressure of nonlinear MHD waves 
and reaches a steady-state characterized by oscillations about 
a new time-averaged equilibrium state;
(2) after turbulent driving is discontinued, the turbulent energy 
dissipates within a few sound crossing 
times of the expanded cloud;
(3) the line-width--size relation is obtained 
by an ensemble of clouds with different free parameters and thereby differing
time-averaged self-gravitational equilibrium states. 
The best consistency with the observational correlation of 
magnetic field strength, turbulent line width, and density 
is achieved by cloud models with $\beta_0 \approx 1$.
We also calculate the spatial power spectra of the turbulent clouds,
and show that significant power is developed on scales 
larger than the scale length $H_0$ of the initial cloud, even if the input
wavelength of turbulence $\lam \approx H_0$, The cloud stratification
and resulting increase of \Alf speed toward the cloud edge allows for a 
transfer of energy to wavelengths significantly larger than $\lam$.  
This explains why the relevant time scale for
turbulent dissipation is the crossing time over the cloud scale rather than
the crossing time over the driving scale.
\end{abstract}

\keywords{ISM: clouds $-$ ISM: magnetic fields $-$ methods: numerical $-$
MHD $-$ turbulence $-$ waves }

\section{Introduction}
Interstellar molecular clouds have nonthermal line widths which
are typically supersonic and increase with the scale of the cloud
size that is measured (e.g., Larson 1981; Myers 1983; Solomon et al. 1987;
Brunt \& Heyer 2002). For a cloud scale $\approx 1$ pc, the 
velocity dispersion $\sigma \approx 1 \kms$, which is several
times greater than the isothermal sound speed $\cs = 0.19 \kms$
for a typical cloud temperature $T = 10 \K$ (Goldsmith \& Langer 1978). 
When detectable through
the Zeeman effect, magnetic field strengths are such that 
their influence is comparable to that of gravity (Crutcher 1999)
and velocity dispersions are correlated with the mean \Alf 
speeds $\bar{V}_A$ such that $\sigma/\bar{V}_A \approx 0.5$ 
(Crutcher 1999; Basu 2000). Furthermore, 
maps of polarized emission from dust imply that the magnetic 
field morphology is relatively 
well-ordered and therefore not dominated by turbulence
(e.g., Schleuning 1998; Matthews \& Wilson 2000; Houde et al. 2004). 
It seems possible that turbulence within molecular clouds comprises 
a spectrum of 
nonlinear magnetohydrodynamic (MHD) waves generated from various
localized sources and from self-gravitational motions.
In contrast to the large-scale structuring by turbulence and magnetic
fields, we note that the embedded dense cores in regions of low-mass
star formation have
characteristically subsonic or transonic turbulence; 
evidence for gravitational collapse in cores also comes in the form of
subsonic infall motions where detectable (see discussion 
and references in Myers 2005).

Although not possessing internal motions as highly 
supersonic as some larger, more distant
molecular clouds, the well-resolved Taurus molecular cloud complex 
(distance $\approx 140$ pc)
reveals a hierarchy of density and size scales that is
characteristic of star-forming regions 
(see Ungerechts \& Thaddeus 1987; Mizuno et al. 1995;
Onishi et al. 1998; Onishi et al. 2002). 
The cloud has a common (turbulent) envelope of molecular column
density $N \approx 1 \times 10^{21} \cms - 2 \times 10^{21} \cms$
(corresponding to visual extinction $A_V \approx 1 - 2$), 
embedded parsec-scale
dark clouds with $A_V \gtrsim 3$ and velocity dispersion 
$\sigma \approx 0.6 \kms \approx 3 \, \cs$, which in turn contain 
dense regions with an average $A_V \approx 5$ that are forming small 
clusters of stars. 
The typical separation of young stellar objects
within the dense regions may be understood by a simple Jeans-like
fragmentation process for an isothermal dense sheet or filament 
(Hartmann 2002). Such an approach ignores the effect of turbulence, 
which is not measured to be large in those regions anyway. However,
the subsonic infall motions imply that a fragmentation process mediated
by magnetic fields and ion-neutral friction may be more appropriate
(Basu \& Ciolek 2004). 

In this paper, we extend a previous set of models (see Kudoh \& Basu
2003, hereafter Paper I) which approximate conditions within the 
molecular cloud envelopes and outer regions of dark clouds.
These regions are expected to be
sufficiently ionized by the far-ultraviolet background starlight
that the ideal MHD limit of flux-freezing is applicable
(McKee 1989). Some key questions about cloud envelopes are as follows.
What is the rate of 
dissipation of the waves and what is the primary dissipation mechanism?
Is continual driving of the turbulence necessary to explain 
their ubiquitous presence in all but the densest regions? 
Furthermore, observed properties of the turbulence such as the
line-width--size relation (e.g., Solomon et al. 1987) and the
correlation between magnetic field strength, line width, and
density (Myers \& Goodman 1988; Basu 2000) deserve an explanation.

Since the late 1990's, a set of three-dimensional models of MHD 
turbulence have been developed which use a {\em local} approach, 
i.e., use periodic boundary conditions (e.g., Stone, Gammie,
\& Ostriker 1998; Mac Low et al. 1998; Mac Low 1999; 
Padoan \& Nordlund 1999; Ostriker, Stone, \& Gammie 2001). 
This means that the modeled regions represent a very small region
deep within a molecular cloud. However, they are driven
with highly supersonic motions characteristic of large-scale 
molecular cloud structure. Although these models are local, 
the turbulent driving
is global, in that large amplitude momentum fluctuations
are input 
at every grid point. The fluctuations are drawn from a Gaussian random
field with a specified root mean squared amplitude and 
power spectrum. Although this method is convenient, a more
likely scenario for molecular cloud turbulence 
is that it is driven from local sources and propagates to fill
the clouds. A main result of  
the periodic-box models is that the decay time 
of the MHD turbulence is comparable to the crossing time over 
the driving scale of the turbulence. For instance, Stone et al.
(1998) find a turbulent dissipation rate $\Gamma_{\rm turb} \approx
\rho_0 \sigma^3/\lam$, where $\rho_0$ is the fixed mean density in the
periodic simulation box, $\sigma$ is the one-dimensional velocity
dispersion, and $\lam$ is the driving scale of turbulence.
A direct application of this result to large molecular clouds has
been questioned by Basu \& Murali (2001), who point out that
dissipation proportional to the crossing time across 
internal driving scales would lead to a CO emission
luminosity $L_{\rm CO}$ far exceeding that observed for clouds 
whose size is much larger than the driving scale.
Alternatively, they point out that the observed scaling of $L_{\rm CO}$
with cloud mass can be explained if the relevant length scale for
dissipation is the largest scale $L$ of any cloud; this also yields
the correct order of magnitude for $L_{\rm CO}$ itself. 
In the context of a periodic box model, this means that
the driving would have to occur only on the largest scales of each cloud.
McKee (1999) also questions extremely fast dissipation, on the 
grounds that:
(1) turbulence should not be so ubiquitous (and present for example
in clouds with
or without embedded OB associations, a possible major source of 
stellar-driven turbulence) if it is decaying so rapidly;
(2) dissipation times of at least a few cloud crossing times 
(as opposed to a free-fall time) may explain the relatively low 
Galactic star formation rate; and (3)
current simulations cannot resolve the 
turbulence within the dense clumps that quickly develop.
Sugimoto, Hanawa, \& Fukuda (2004) have supported
the last point with second-order accurate 
three-dimensional numerical studies of 
\Alf wave propagation, finding that $\simeq 30$ grid points
per wavelength are necessary to reduce dissipation to 1\% per
wavelength propagated. Such a high resolution remains prohibitive
for three-dimensional simulations that seek to model large-scale
cloud dynamics and dense core formation.
Cho \& Lazarian (2003) have also questioned a common interpretation
of three-dimensional periodic box models, i.e., that the
fast dissipation is due to the coupling of \Alf modes to the
more dissipative fast MHD and slow MHD modes (e.g., Stone et al. 1998). 
Through modal analysis of their three-dimensional simulations, 
they find that the coupling of modes 
is rather weak in the inertial range of turbulence and that 
dissipation of \Alf waves is dominated by a nonlinear cascade.
Furthermore, Cho, Lazarian, \& Vishniac (2002) have found that
in the case of incompressible MHD turbulence, a directionality in
the flux of energy (as may be expected with localized driving
sources) leads to a significantly lower dissipation rate.

In addition to the 
detailed analysis of individual wave modes as 
described above, new approaches to the numerical study of 
MHD turbulence include
extremely high resolution models of the {\em global} structure of  
molecular clouds
(Kudoh \& Basu 2003; Folini, Heyvaerts, \& Walder 2004).
These one-dimensional models employ a numerical resolution 
that remains beyond the reach of three-dimensional simulations.
Kudoh \& Basu (2003, hereafter Paper I) have modeled the 
global structure of a one-dimensional cold cloud that is
embedded in a hotter external medium. The main advantage over
a local (periodic box) approach is the inclusion of the effects 
of cloud stratification and a cloud edge. Furthermore, the turbulent
driving is localized and the propagation of disturbances 
within the cloud as well as their leakage from the cloud can be
studied. Our model in Paper I had 50 numerical cells within
the scale length $H_0$ of the initial state, and 70 cells
within the wavelength $\lam$ associated with turbulent driving.
Results included the generation of significant density structure
and shock fronts within the cloud; however, most of the wave
energy remained in transverse, rather than longitudinal motions
(in qualitative agreement with the results of Cho \& Lazarian
2003) and the time-averaged density profile of the cloud
(which may be similar to an observed spatial average of many 
layers along the line-of-sight) was relatively smooth. 
The localized driving led to turbulence which quickly
filled the cloud, causing cloud expansion and 
large-scale oscillations in the outer regions; all this is 
aided by a clear transfer of energy to the largest
scales of the expanded cloud, due to the stratification
and consequent increase of \Alf speed toward the cloud edge.
The work of Folini et al. (2004), who model a very similar system
with similar turbulent driving, confirm the existence of
highly disturbed density structure, oscillatory motions, and
a dominance of wave energy in transverse modes. 
They focus most extensively on
the density structuring and demonstrate that progressively finer scale 
structuring and enhanced cloud support is revealed by higher resolution
simulations; for typical dimensional values of physical parameters, 
their numerical resolution is 0.001 pc, as in our models in Paper I.
In Paper I, we also considered many global properties of the clouds.
We found that 
random motions were greatest in the outer, low-density
regions of the clouds in a way that the velocity dispersion $\sigma$ 
and cloud scale $L$ satisfied the observed line-width--size relation.
Our model also showed that even when turbulence decays away due to lack of
driving, the large-scale oscillations are the longest lived component.
These results can help to explain the observations of
apparent oscillations of B68 (Lada et al.
2003), a dark cloud in which nonthermal motions are small. 
We also showed that the overall dissipation
of turbulent energy (after driving is discontinued) 
is proportional to the crossing time across the expanded 
cloud, rather than the
crossing time for the driving wavelength $\lam$. This result points
to a way out of the problems with internal driving pointed out by
Basu \& Murali (2001). Stratification leads 
to turbulence at the largest scales, so that the cloud
scale $L$ is more relevant to turbulent dissipation than the internal
driving scale $\lam$, as needed on empirical grounds. 
We quantify the effect of turbulence on large scales further in 
this paper by considering the power spectrum of fluctuations in the
turbulent clouds.

This paper expands upon the study of Paper I and summarizes 
the results of a parameter study. The numerical model is presented
in \S\ 2. The results of the simulations are described 
in \S\ 3. A summary and discussion are presented in \S\ 4. 

\section{The Numerical Model}

\subsection{Overall Approach}

The basic model is the same as presented in Paper I.
We consider a molecular cloud that is threaded by a
large-scale magnetic field and assume ideal MHD motions. 
We assume a driving force near 
the midplane of the cloud and follow the dynamical evolution 
of the vertical structure of the cloud.
Our model can be characterized as 1.5-dimensional, since we calculate 
variations in only one direction but allow vector quantities to 
have both a longitudinal and a transverse component (see details in
\S\ \ref{basic}).
We also assume isothermality for each Lagrangian fluid element.
This means that the
temperature does not change in time for each fluid element
as it moves through Eulerian space. Since we have a two-temperature
model (cloud and intercloud medium), this is not the same as a 
purely isothermal model.

\subsection{Basic Equations and Physical Variables}
\label{basic}

We use local Cartesian coordinates $(x,y,z)$ on the molecular cloud, 
where we set $z$ to be the direction of the large-scale magnetic field. 
According to the symmetry of the 1.5-dimensional approximation, we set
\begin{equation}
\frac{\partial}{\partial x}=\frac{\partial}{\partial y}=0.
\end{equation}
The above symmetry and the divergence-free condition on the magnetic field
imply
\begin{equation}
B_{z} = {\rm constant},
\label{eq:Bz}
\end{equation}
where $B_z$ is the $z$-component of the magnetic field that 
threads the molecular cloud. Moreover, from the assumption of linear 
polarization of the waves, we can set
\begin{equation}
v_x=B_x=0
\end{equation}
without loss of generality, where $v_x$ and $B_x$ are the $x$-components 
of the velocity and magnetic field, respectively.

Under this assumption, we solve the MHD equations numerically.
In the equations, $t$ is the time, 
$G$ is the gravitational constant, 
$k$ is Boltzmann's constant, 
$m$ is the mean molecular mass, 
$\rho$ is the density, $P$ is the pressure, $T$ is the temperature, 
$g_z$ is the $z$-component of the gravitational field, 
$v_z$ is the $z$-component of velocity, 
$v_y$ is the $y$-component of velocity, and 
$B_y$ is the $y$-component of the magnetic field. 
In the equations, the energy equation is assumed to be
\begin{equation}
   {\partial T \over \partial t}
    + v_{z} {\partial T \over \partial z}
    = 0,
\label{eq:tlag}
\end{equation}
which quantifies the assumption of 
isothermality for each Lagrangian fluid element.

\subsection{Initial Conditions}

As an initial condition, we assume hydrostatic equilibrium
of a self-gravitating one-dimensional cloud.
The hydrostatic equilibrium is calculated from the equations
\begin{equation}
\frac{1}{\rho}\frac{d P}{d z} = g_z,
\label{eq:hdsr}
\end{equation}
\begin{equation}
\frac{d g_z}{d z} = -4\pi G \rho,
\label{eq:hdsg}
\end{equation}
and
\begin{equation}
P = \rho \frac{kT}{m},
\label{eq:hdsp}
\end{equation}
subject to the boundary conditions
\begin{equation}
g_z(z=0)=0, 
\end{equation}
\begin{equation}
\rho(z=0)=\rho_0,
\end{equation}
and
\begin{equation}
P(z=0)= \rho_0 \frac{kT_0}{m},
\end{equation}
where $\rho_0$ and $T_0$ are the initial density and temperature 
at $z=0$, respectively.

In order to solve the above equations, we need to assume an 
initial temperature distribution.
If the temperature is uniform throughout the region,
we have the following analytic solution $\rho_{\rm S}$ found by 
Spitzer (1942):
\begin{equation}
\rho_{\rm S} (z)=\rho_0 \, {\rm sech}^2 (z/H_0),
\end{equation}
where
\begin{equation}
H_0=\frac{c_{s0}}{\sqrt{2 \pi G\rho_0}}
\end{equation}
is the scale length, and 
\begin{equation}
c_{s0}=\sqrt{ \frac{k T_0}{m} }
\end{equation}
is the isothermal sound speed for temperature $T_0$.

However, an isothermal molecular cloud is usually surrounded by warm 
or hot material, such as neutral hydrogen or ionized gas.  
Therefore, we assume the initial temperature distribution 
to be
\begin{equation}
T(z)=T_0+\frac{1}{2}(T_c-T_0)\left[ 1+\tanh \left(\frac{|z|-z_c}{z_d} \right)\right],
\end{equation}
where we take $T_c=100\,T_0$, $z_c=3H_0$, and $z_d=0.2H_0$ throughout the paper.
This distribution shows that the temperature is uniform and equal to $T_0$
in the region of $0 \leq z < z_c=3H_0$ and smoothly increases to
another uniform value $T_c=100T_0$ at $z \simeq z_c=3H_0$.
By using this temperature distribution, we can solve the 
ordinary differential equations (\ref{eq:hdsr})-(\ref{eq:hdsp}) 
numerically. 

We also assume the following initial conditions:
\begin{equation}
v_z(z)=v_y(z)=0, 
\end{equation}
\begin{equation}
B_y(z)=0, 
\end{equation}
\begin{equation}
B_z(z)=B_{0}, 
\end{equation}
where $B_{0}$ is a constant. According to equation (\ref{eq:Bz}),
$B_z$ is spatially uniform and independent of time throughout 
the calculations. 

\subsection{Driving Force}

We introduce a perturbation into the initially hydrostatic cloud
by adding a driving force, $F(z,t)$,
into the $y$-component of the momentum equation 
as follows:
\begin{equation}
   \rho ({\partial v_{y} \over \partial t}
    + v_{z} {\partial v_{y} \over \partial z})
    = {1 \over 4 \pi} B_{z} {\partial B_{y} \over \partial z}
      + F(z,t).
\label{eq:driving}
\end{equation}

In this paper, we use two forms of the driving force.
The first one, which is the same as in Paper I, is the sinusoidal 
driving force defined by
\begin{equation}
F(z,t) = \left\{
\begin{array}{ll}
\rho a_d (\frac{t}{10t_0}) \sin(2\pi \nu_0 t) \exp[-(\frac{z}{z_a})^2] & (t<10t_0) \\
\rho a_d \sin(2\pi \nu_0 t) \exp[-(\frac{z}{z_a})^2]  & (10t_0 \leq t \leq  40t_0) \\
0  & (t> 40 t_0), 
\end{array}
\right.
\label{eq:driving2}
\end{equation}
and 
\begin{equation}
t_0=H_0/c_{s0}.
\end{equation}
In this form, $a_d$ is the amplitude of the induced acceleration,
$\nu_0$ is the frequency of the driving force,
and $z_a$ represents the region in which we input the driving force.

The other form we use is the random driving force:
\begin{equation}
F(z,t) = \left\{
\begin{array}{ll}
\rho a_d (\frac{t}{10t_0}) {\rm ran}(t) \exp[-(\frac{z}{z_a})^2] & (t<10t_0) \\
\rho a_d {\rm ran}(t) \exp[-(\frac{z}{z_a})^2]  & (10t_0 \leq t \leq  40t_0) \\
0  & (t> 40 t_0), 
\end{array}
\right.
\label{eq:drivingr}
\end{equation}
where ${\rm ran}(t)$ is a random number between -1 and 1 which 
is determined at each time step.

These equations show that we input the driving force near 
the midplane of the cloud, and increase the maximum driving force 
linearly with time until $t=10t_0$, and maintain it to be 
constant during $10t_0 \leq t \leq  40t_0$. 
After $t=40t_0$, we terminate the driving force.

\subsection{Boundary Conditions and Numerical Technique}

We used a mirror symmetric boundary condition at $z=0$ and 
a free boundary at $z = z_{\rm out}$, the outer boundary of 
the calculation.
In order to remove the reflection of waves
at the outer boundary, we set $z_{\rm out}$ to be a large value, 
and use nonuniform grid spacing for large $z$. 
For $z \leq 80H_0$, the grid spacing is $\Delta z_i = 0.02 H_0$;
most of the $4000+$ points in a typical simulation are concentrated
in this region. Such high resolution is feasible in a 1.5-dimensional
simulation and ensures that the dissipation rate is determined by physical 
effects like a nonlinear cascade and nonlinear steepening rather than by
direct numerical diffusion (see details in Paper I). 

In order to solve the equations numerically, we use the CIP method
(e.g., Yabe, Xiao \& Utsumi 2001) and the MOCCT method 
(Stone \& Norman 1992).
The combination of the CIP and MOCCT methods is summarized in 
Kudoh, Matsumoto \& Shibata (1999). The CIP method is a useful 
method to solve an advection equation such as equation 
(\ref{eq:tlag}) accurately, and is also applicable to 
advection terms of hydrodynamic equations.
Due to the mirror-symmetric boundary condition at $z=0$,
the Poisson equation can be simply integrated from 
the midplane of the cloud.

\subsection{Numerical Parameters}

A natural set of fundamental units for this problem are
$c_{s0}$, $H_0$, and $\rho_0$. These yield a time unit
$t_0=H_0/c_{s0}$.
The initial magnetic field strength introduces one dimensionless
free parameter, i.e.
\begin{equation}
\beta_0 \equiv \frac{8\pi P_0}{B_0^2}=\frac{8\pi\rho_0 c_{s0}^2}{B_0^2},
\end{equation}
which is the initial ratio of gas to magnetic pressure at $z=0$.

In this cloud, $\beta_0$ is related to the mass-to-flux ratio.
For Spitzer's self-gravitating cloud, the mass-to-flux ratio normalized
to the critical value is
\begin{equation}
\mu_{\rm S} \equiv 2 \pi G^{1/2} \frac{\Sigma_{\rm S}}{B_0},
\end{equation}
where 
\begin{equation}
\Sigma_{\rm S}=
\int_{-\infty} ^\infty \rho_S \ dz =
2 \rho_0 H_0
\end{equation}
is the column density of the Spitzer's self-gravitating cloud.
Therefore,
\begin{equation}
\beta_0=\mu_{\rm S}^2.
\label{eq:bemu}
\end{equation}
The column density of the cloud we used in this paper is almost 
equal to that of Spitzer's cloud.
If we define the column density of the cloud, $\Sigma$, 
as the integral of density within  $-z_c <z< z_c$, then
\begin{equation}
\Sigma = \int_{-z_c} ^{z_c} \rho (t=0) dz \simeq 0.988 \, \Sigma_{\rm S}.
\end{equation}
This means that we can use the value of $\mu_{\rm S}$ an excellent 
approximation to the dimensionless mass-to-flux ratio of the model cloud.

The driving force introduces three more free parameters:
$\tilde{a}_d=a_d (H_0/c_{s0}^2)$, the dimensionless amplitude
of the acceleration due to driving; $\tilde{\nu}_0=\nu_0 t_0$,
the dimensionless frequency of driving; 
$\tilde{z}_a=z_a/H_0$, the dimensionless scale of the
driving region. For simplicity, we take $\tilde{z}_a=0.1$ 
throughout this paper.

Dimensional values of all quantities can be found through
a choice of $T_0$ and $\rho_0$, along with the values
of the dimensionless free parameters.
For example, if $T_0=10 \K$ and $n_0=\rho_0/m= 10^4 \cmc$,
then $c_{s0}=0.2 \kms$, $H_0=0.05 \pc$, 
$N_S=\Sigma_S/m=3 \times 10^{21} \cms$,
$t_0=2.5 \times 10^5$ yr, and $B_0=20 \muG$ if 
$\beta_0=1$.

\subsection{Oscillatory Motions}

Due to the nonlinearity and inherent randomness in the model, a 
non-negligible mean transverse motion of the cloud can be 
set up after a few cycles of turbulent driving. In order to exclude the
energy of the mean transverse cloud motion from our analysis of turbulent
motions, we introduce (as in Paper I) the oscillating component of 
the $y$-velocity,
\begin{equation}
v^\prime_y=v_y-\langle v_m(t)\rangle,
\label{eq:vyosc}
\end{equation}
where
\begin{equation}
v_m(t) = \frac{\int_0^{z_f(t)} \rho v_y dz}{\int_0^{z_f(t)} \rho dz} \ .
\end{equation}
In this equation, $v_m$ is the mean $y-$velocity of the cloud and 
$z_f(t)$ is the full mass position of the cloud, which
is defined by
\begin{equation}
\int_0^{z_f(t)} \rho dz = 0.998 \frac{\Sigma_{\rm S}}{2}.
\end{equation}
This means that $z_f(t)$ corresponds to the position of the 
Lagrangian fluid element
which is initially located at $\simeq z_c$, the initial position 
of the transition region of the temperature.

In equation (\ref{eq:vyosc}), we take a time average of $v_m$ for 
each cycle of the sinusoidal period ($T_0=1/\nu_0$) of the driving 
force in order to remove the oscillation caused by the driving force.
The time average, $\langle v_m\rangle$, is defined
in the following manner. 
For example,  $\langle v_m\rangle$ 
between $nT_0$ and $(n+1)T_0$ is calculated as 
\begin{equation}
\langle v_m\rangle(t=nT_0-(n+1)T_0) = \frac{1}{T_0} 
\int_{nT_0}^{(n+1)T_0} v_{\rm m}(t^\prime) dt^\prime,
\end{equation}
where $n$ is an integer. 
According to the definition, $\langle v_m\rangle$ has the same value 
between $nT_0$ and $(n+1)T_0$. When the driving force is random, 
there is really no period. In this case, we choose $T_0=t_0$ for convenience.

\section{Results}

Table 1 summarizes the parameter study we have performed.
The first 5 models with different $\tilde{a}_d$ are the same 
as presented in Paper I. In this paper, we primarily discuss the results 
of varying $\tilde{\nu}_0$ and $\beta_0$, as well as the effect of a 
random driving force. We describe in detail a subset of the full range 
of models listed in Table 1.

\subsection{Cloud Dynamics}
\label{sec:clouddyn}

Figure 1 shows the time evolution of models with different frequencies
$\nutil$.
The left panel shows the model with $\tilde{\nu}_0=0.5$, 
the middle panel shows the model with $\tilde{\nu}_0=1$,
and the right panel shows the model with $\tilde{\nu}_0=2$;
all models have $\tilde{a}_d=30$ and $\beta_0=1$.
For each model, the upper panels show the time evolution of the density.
The density profile at various times are stacked with time
increasing upward in uniform increments of $0.2t_0$.
The lower panels show the time evolution of $v'_y$
at $z=0$, where $v'_y$ is the oscillating component 
of the $y$-velocity that we defined in the previous section.
As we have shown in Paper I, each density evolution plot shows many
shock waves propagating in the cloud and significant upward
and downward motions of the outer portion of the cloud
during the period of constant energy input ($t=10t_0-40t_0$).
Figure 1 shows that the cloud undergoes more expansion and the amplitude
of $v'_y$ becomes greater when the frequency is lower.
We return to this point after examining the effect of $\beta_0$.

Figure 2 presents the same physical quantities as Figure 1, but for
models with different values of $\beta_0$.
The left panel shows the model with $\beta_0=0.16$, 
the middle panel shows the model with $\beta_0=0.25$,
and the right panel shows the model with $\beta_0=4$;
all models have $\tilde{a}_d=30$ and $\tilde{\nu}_0=1$.
When $\beta_0=0.16$, the cloud hardly expands, showing no shock waves or
large up-down motions. The velocity amplitude $v'_y$ at $z=0$ is 
quite small in comparison to other models, and typically subsonic. 
This model has the strongest magnetic field for a given density, and
corresponds to the most magnetically subcritical case. 
When $\beta_0=0.25$, the cloud shows large 
oscillations, although there are not as many shock waves as in 
the standard model with $\beta_0=1$ (middle panel of Fig. 1).
The case of $\beta_0=4$ shows the greatest velocity amplitude
$v'_y$ at the midplane, and many shock waves. This is consistent
with the trend set by other models that a weaker magnetic field
leads to greater disturbances at the midplane for a given driving
amplitude $\atil$. The $\beta_0=4$ model is significantly magnetically
supercritical ($\mu_{\rm S} = 2$). However, in contrast to
the trend established for increasing $\beta_0$
by the other models, the cloud does not expand nearly as much as 
in the $\beta_0=0.25$ or $\beta_0=1$ models.

The differences in the level of cloud expansion for the models with
differing $\nutil$ and/or $\beta_0$ can be understood in part by considering 
the different wavelengths of the input driving. The input wavelength 
$\lambda_0$ is estimated from the \Alf velocity at the midplane
$V_{A0}$ and the period $T_0$ of the driving force:
\begin{equation}
\lambda_0 \approx V_{A0} T_0 = \frac{B_0}{\sqrt{4\pi\rho_0}} \frac{1}{\nu_0}
        = \frac{\sqrt{2} H_0}{\tilde{\nu}_0 \sqrt{\beta_0}} \ .
\end{equation}
For the models presented in Figure 1 with $\nutil = 0.5, 1, 2$, the corresponding
input wavelengths are $\lam = 2.83H_0, 1.41H_0, 0.71H_0$, respectively.
The model with $\nutil=0.5$ has an input wavelength closest to that of the
initial cloud size ($\approx 3.0H_0$), and this leads to the greatest 
effect of cloud expansion due to the pressure of the nonlinear waves.
This correlation is borne out further by examining the models of different
$\beta_0$ presented in Figure 2.
When $\beta_0=0.16$ and $\tilde{\nu}_0=1$, in addition to the low
amplitude fluctuations $v'_y$ due to the very strong magnetic field,
we also have $\lambda_0 \approx 3.5H_0$, which is larger than the initial cloud size.
Hence, the waves are not effective at providing an internal pressure.
When $\beta_0=0.25$ and 
$\tilde{\nu}_0=1$ we find $\lambda_0 \approx 2.83H_0$, which is 
approximately the same as the initial cloud size. 
On the other hand, when $\beta_0=4$ and 
$\tilde{\nu}_0=1$ we get $\lambda_0 \approx 0.71H_0$, which is significantly
smaller than the cloud size. Even though there are large velocity fluctuations
$v'_y$ at the midplane in this model due to the relatively weak magnetic
field, the overall expansion of the cloud is inhibited by the relatively
small ratio $\lam/H_0$.
Our overall results indicate that 
the cloud suffers less expansion when the 
driving wavelength is larger than the initial cloud size, and
the cloud is lifted up the most when the driving wavelength
is approximately the same as the initial cloud size.

Figure 3 and Figure 4 show  other models 
which differ from the standard model in the values of 
{\em two} separate parameters.
The left panel of Figure 3 shows the model with $\beta_0=0.25$ and 
$\tilde{\nu}_0=0.5$, while the right panel shows the model with $\beta_0=4$ and 
$\tilde{\nu}_0=0.5$; both models have $\tilde{a}_d=30$.
The driving wavelength $\lambda_0 \approx 5.66H_0$ for the model
in the left panel, and $\lambda_0 \approx 1.41H_0$ for the model
in the right panel.
These two panels are the low frequency counterparts of the
middle and right panels of Figure 2, respectively. 
Figure 3 also shows that the cloud suffers a lesser expansion
when the driving wavelength is larger than the cloud size.

The left panel of Figure 4 shows the model with $\beta_0=0.25$ 
and $\tilde{\nu}_0=2$, and the right panel shows the model with
$\beta_0=4$ and $\tilde{\nu}_0=2$; both models have $\tilde{a}_d=30$.
The driving wavelength is $\lambda_0 \approx 2.83H_0$ for the model
in the left panel, and $\lambda_0 \approx 0.354H_0$ for the 
model in the right panel.
These two models are the high frequency counterparts of the
models in Figure 3. In contrast to Figure 3, the model with $\beta_0=0.25$
shows greater expansion because the wavelength is comparable to
the initial cloud size, while the model with $\beta_0=4$ shows
very little expansion since the driving wavelength is much smaller than
than the initial cloud size.

In Figure 5, we show the results of the models with a random driving 
force. The models have different values of $\beta_0$ and the same 
dimensionless amplitude 
$\tilde{a}_d=5000$. Since the random driving contains various 
frequencies, the time evolution of the density is not significantly 
different from model to model. Each model has a component of the 
driving at the wavelength (which is a function of $\beta_0$) that is 
comparable to the cloud size. However, the amplitudes 
of $v'_y$ are a little different in each model. When the magnetic 
field is strong ($\beta_0=0.25$), the amplitude is smaller, and 
vice versa. This result is consistent with the results presented 
in Figure 2.

To conclude this section, we reemphasize that the wavelength
of the driving force is an important parameter for the cloud 
evolution. The driving wavelength which is comparable to 
the cloud size has the greatest effect in expanding the cloud.
Thus, the expansion cannot be attributed primarily to the commonly 
used notion of ``turbulent pressure'', which requires that the 
wavelength of disturbances be much smaller than the typical scale
length of the cloud (see Bonnazola et al. 1992; Martin, Heyvaerts,
\& Priest 1998).

\subsection{Energy Dissipation Rate}

Figure 6 shows the time evolution of energies in the cloud 
for the case of random driving, $\tilde{a}_d=5000$ and $\beta_0=1$.
Each energy is calculated in the following manner:
kinetic energy of the $z$-component of velocity,
\begin{equation}
E_{kz} (t) = \int_0^{z_f(t)} \frac{1}{2} \rho v_z^2 \, dz;
\label{eq:ekz}
\end{equation}
kinetic energy of the $y$-component of velocity,
\begin{equation}
E_{ky} (t) = \int_0^{z_f(t)} \frac{1}{2} \rho v_y^2 \, dz 
- \, E_{km},
\label{eq:eky}
\end{equation}
where $E_{km}$ is the kinetic energy of the mean motion of the cloud, i.e.,
\begin{equation}
E_{km}(t)= \frac{1}{2} \langle v_m\rangle^2 \int_0^{z_f(t)} \rho \, dz;
\end{equation}
magnetic energy of the $y$-component of the magnetic field,
\begin{equation}
E_{m} (t) = \int_0^{z_f(t)} \frac{B_y^2}{8\pi}  \, dz;
\label{eq:em}
\end{equation}
the sum of the above terms,
\begin{equation}
E_{T} (t) = E_{kz} (t) + E_{ky} (t) + E_{m} (t).
\label{eq:et}
\end{equation}
In the above equations, the integration was done from 0 to $z_f(t)$ 
because we are interested in the energies of the cold material.
Strictly speaking, the total energy in each case is twice the value
we calculate due to the mirror symmetric boundary condition at $z=0$.
Equation (\ref{eq:eky}) shows that the mean kinetic energy of the 
cloud is subtracted from the kinetic energy of the the $y$-component of the velocity,
so that $E_{ky}$ is the kinetic energy of the oscillatory motions.

In Figure 6, we examine the evolution of various energies for the
random
driving force model with $\tilde{a}_d=5000$ and $\beta_0=1$.
The thick solid line shows $E_{ky}$, the thin solid 
line shows $E_{m}$, the dotted line shows $E_{kz}$, and 
the dashed-dotted 
line shows $E_{T}$. The values of each energy are smoothed out 
over a time $t=t_0$ to remove small oscillations. Figure 6 is similar 
to the Figure 8 in Paper I, which shows the evolution of these
quantities in the standard model with sinusoidal driving force.
Among the energies, $E_{ky}$ and $E_{m}$ are comparable to each
other, but $E_{kz}$ is significantly smaller, 
i.e., there is approximate equipartition in transverse kinetic energy and
magnetic fields, and there is much less energy in longitudinal modes.
Longitudinal modes can be generated by the gradient of magnetic
field pressure in the $z$-direction.
However,
the dominance of the transverse modes implies that the primary 
dissipation mechanisms are the nonlinear cascade and nonlinear steepening
effects that transfer energy to progressively smaller scales
until it is dissipated at the grid scale (see more extensive discussion
in Paper I).
After the driving force is terminated at $t=40t_0$, the energies
decrease almost exponentially.
The $e$-folding time is $t_d \simeq 10t_0$, which is approximately a 
sound crossing time across the cloud size as expanded by the
turbulent pressure. The turbulent energy is nearly completely dissipated
within a few of these crossing times. 
All of this is consistent with the results
for sinusoidal driving presented in Paper I.

Figure 7 shows the time evolution of $E_{T}$ for different 
values of $\tilde{\nu}_0$ with fixed values of $\beta_0=1$ and 
$\tilde{a}_d=30$.
Figure 8 also shows the time evolution of $E_{T}$ for different 
values for $\beta_0$ but with fixed values $\tilde{\nu}_0=1$ and 
$\tilde{a}_d=30$.
The energy decreasing time appears to be somewhat
lesser when the driving frequency is higher.
The $e$-folding time of dissipation for each parameter is listed in Table 1.
It appears to have a weak correlation with the cloud size. 
The smallest $e$-folding times ($t_d <4.0t_0$) occur when 
the wave length $\lambda_0$ is larger than the initial cloud size 
$\sim 3H_0$. In this case, most of the input energy escapes
from the cloud without affecting it much.
Overall, our results show that $t_d \approx 10 t_0$ is a good approximation 
for most clouds, i.e., that 
$t_d$ does not depend very sensitively on most parameters. 

\subsection{Spatial Power Spectrum}

Figure 9 shows the spatial power spectra of $B_y$ and $v_y$ for
a model with $\tilde{a}_d=30$, $\tilde{\nu}_0=1$, and $\beta_0=1$.
Both are plotted against the dimensionless quantity $k H_0$, where
$k = 1/\ell$ and $\ell$ is a length scale in the $z$-direction.
The power spectra were taken at $t=30t_0$, in the midst of the
period of steady turbulent driving, over the spatial range 
$z=0-60H_0$.
There is a weak indication of a local peak in the power spectrum of 
$B_y$ around $k = 1/\lam \approx 0.70/H_0$, which corresponds to 
the driving scale $\lam \approx 1.41 H_0$. 
However, the main result here is that there is significant 
power on scales larger than the driving scale ($k <  0.70/H_0 $)
in the spectra of both $B_y$ and $v_y$. 
The dotted line shows the dependence $k^{-5/3}$ for comparison; it is
the Kolomogorov power spectrum for one-dimensional incompressible
hydrodynamic motions
(e.g., Elmegreen \& Scalo 2004). Our simulations of compressible MHD 
motions shows both spectra are similar but 
have slightly steeper than $k^{-5/3}$ slopes on small scales($k > H_0^{-1}$).
Figure 10 shows the same power spectra for 
the random perturbation model with $\tilde{a}_d=1000$, and $\beta_0=1$.
These power spectra confirm the basic result of the 
sinusoidal models that there is significant 
power on large scales ($k < H_0^{-1}$) and that the 
slopes are steeper than $k^{-5/3}$ on small scales($k > H_0^{-1}$).

The power spectra show that there is significant power 
on the largest scales in the cloud and that the power
spectrum naturally evolves to the form $k^{-p}$, where $p$
is a positive number, even if driving occurs within the 
cloud on a sub-cloud scale.
This large-scale power comes 
from the wave propagation in the gravitationally stratified 
cloud. When a wave propagates into the low density region
with greater \Alf speed, 
the wavelength becomes larger. The waves of larger wavelengths 
also generally have less dissipation. These principles lead to
significant power on larger scales in a gravitationally 
stratified cloud.

\subsection{Correlations between Velocities and Sizes}

Figure 11 plots the time averaged velocity dispersions 
$\langle \sigma^2 \rangle_t^{1/2}$ of 
Lagrangian fluid elements within various model clouds
as a function of $\langle z \rangle_t$, where
\begin{equation}
\sigma=\sqrt{\onehalf [v_z^2+(v^\prime_y)^2] + c_s^2},
\end{equation}
and $\langle z \rangle_t$ is the time-averaged height 
of each Lagrangian fluid element.
The open circles ($\sigma_F$, $z_F$ in Table 1)
correspond to Lagrangian fluid elements
whose initial positions are located at $z=2.51 H_0$, 
which is close to the edge of the cold cloud.
The filled circles ($\sigma_H$, $z_H$ in Table 1)
correspond to Lagrangian fluid elements
whose initial positions are located at $z=0.61 H_0$, 
which is approximately the half-mass position of the cold cloud.
An open and filled circle is plotted for each model in our complete
parameter study; some basic properties of each model are 
summarized in Table 1\footnote[1]{We note that the 
velocity dispersions listed in Table 1 
of Paper I were incorrect, although 
the Figure 13 of Paper I was plotted using the correct
values of velocity dispersion. This Table contains the correct 
velocity dispersions for all models presented here and in Paper I.}.
The dotted line shows  
\begin{equation}
\langle \sigma^2 \rangle_t^{1/2} \propto \langle z \rangle_t^{0.5}.
\end{equation}
This figure shows that the model clouds are in a time-averaged 
self-gravitational equilibrium state under the influence of
turbulent pressure (see Paper I). The varying parameters can affect the 
position of the circles, but all model clouds obey the basic correlation
shown by the dotted line.
This is consistent with the observational 
line-width--size relation of molecular clouds 
(Larson 1981; Myers 1983; Solomon et al. 1987).

Figure 12 shows the correlation between the velocity dispersion
and mean Alfv\'en velocity of the cloud $\bar{V}_A$ at 
the mean position $\langle z \rangle_t$ of the Lagrangian elements. 
The mean Alfv\'en velocity is defined by
\begin{equation}
\bar{V}_A \equiv \frac{B_0}{\sqrt{4\pi\bar{\rho}}},
\label{eq:mva}
\end{equation}
where 
\begin{equation}
\bar{\rho}=\frac{\Sigma}{2 \langle z \rangle_t}
\label{eq:mde}
\end{equation}
is the mean density and $\Sigma$ is the column density for each 
Lagrangian element.
The dotted line shows  
\begin{equation}
\langle \sigma^2 \rangle_t^{1/2} \propto \bar{V}_A.
\label{eq:lwva}
\end{equation}
Results from model clouds with the same $\beta_0$ are circled and marked.
They show that the velocity dispersions have a good correlation with 
the mean Alfv\'en velocity if each cloud has a similar value
of $\beta_0$, i.e. mass-to-flux ratio.
An analysis of the magnetic field measurements compiled by Crutcher (1999)
shows that there is an excellent correlation
between the line-of-sight component of the large-scale magnetic field 
$B_{\rm los}$ and $\sigma \rho^{1/2}$ for observed clouds of widely 
varying values of $\rho$, $B_{\rm los}$, and $\sigma$ (Basu 2000). 
This is essentially 
the same correlation as equation (\ref{eq:lwva}); the observational data
actually imply an average value $\sigma/\bar{V}_A \approx 0.5$ (see
Basu 2005 for discussion). Our models are consistent with this ratio if
$\beta_0 \approx 1$, i.e the mass-to-flux ratio is essentially equal
to the critical value. A comparison of observed values of $B_{\rm los}$
and column density also reveals that the mass-to-flux ratios are
always very close to the critical value (Crutcher 1999; Shu et al. 1999;
Crutcher 2004; Basu 2005).

\section{Summary and Discussion}

We have performed 1.5-dimensional numerical simulations of nonlinear 
MHD waves in a gravitationally stratified molecular cloud that 
is bounded by a hot and tenuous external medium. 
Using the same basic model as presented in Paper I (Kudoh \& Basu 2003), 
we have carried out a parameter survey
by varying the frequency of the driving force and the magnetic field
strength of the cloud. We found that the key parameter for 
the evolution of the cloud is the Alfv\'en wavelength of
the driving force. If the wavelength is larger than
the size of the cloud, the cloud is affected less by 
the waves. The wavelength that is the same order of the
cloud size is the most effective in expanding the cloud.
This means that turbulent expansion is different than
the usual notion of expansion due to a ``turbulent pressure'' in 
which the wavelengths need to be much smaller than the cloud size.
We also studied the effect of a driving force that varies
randomly with time and found that significant cloud expansion occurs since 
there is always a component of the driving that is
at a favorable frequency for expanding the cloud. 
The evolution of our model clouds reveal the following general
features:
(1) Under the influence of a driving source of Alfv\'enic disturbances,  
a cloud shows significant upward and downward motions, with an
oscillation time scale 
that is comparable to the cloud crossing time;
(2) After driving is discontinued, the turbulent energy dissipates
exponentially with time, with an $e$-folding time essentially equal to
the crossing time 
across the expanded cloud rather than the crossing time of the
driving scale; (3)
The power spectra show that significant power is generated on 
scales larger than that of the turbulent driving ($k < H_0^{-1}$);
(4)
The line-width--size relation is obtained by an ensemble of 
clouds with different physical parameters which are individually in a 
time-averaged self-gravitational equilibrium state. The largest
amplitude random motions occur in the outer low density regions of 
a stratified cloud. The magnetic
field data is best fit by models with $\beta_0 \approx 1$, which implies
that the mass-to-flux ratio is very close to the critical value
for collapse.


In the case of the random driving force, we require a much larger driving
amplitude $\tilde{a}_d$ than for the
sinusoidal case in order to get a comparable expansion of the cloud
(see Table 1). This is due to the multiplication by a random number
in the range $[-1,1]$ at each time step (see eq. [\ref{eq:drivingr}]) in the
random case.
The resulting driving efficiency is quite low since the driving
force $F(z,t)$ is often out of phase with $v_y$, resulting in a lesser
amount of work done in the random case than in the sinusoidal case for
a given value of $\tilde{a}_d$.
A physically meaningful comparison of the energetics of the random 
and sinusoidal cases can be done by calculating the energy flux of 
\Alf waves emerging from just outside the region of the artificial driving. 
We define the \Alf energy flux in the $z$-direction to be
\begin{equation}
{\cal F}_A=-\frac{B_z B_y}{4\pi} (v_y-\langle v_m\rangle) ,
\end{equation}
and perform the calculation at $z=z_a=0.1 H_0$.
We find that the time-averaged energy flux of a random driving model 
with parameters $\tilde{a}_d=5000$ and $\beta_0=1$ is comparable to 
(actually about twice as large as) that of a sinusoidal driving model
with parameters $\tilde{a}_d=30$, $\beta_0=1$, and $\nu_0=1$. 

For the random driving case ($\tilde{a}_d=5000$, $\beta_0=1$), 
the time-averaged energy flux measured near the surface of 
the cloud (at the full-mass position) is 
about 20\% of that at $z=z_a$, so that about 80\% of the input 
energy is dissipated in the cloud. For the sinusoidal driving case 
($\tilde{a}_d=30$, $\beta_0=1$, $\nu_0=1$), the energy flux near the surface 
is about 30\% of that at $z=z_a$. 
The small difference may arise because the random model contains many 
high frequency waves which dissipate quickly.
In each case, the dissipated energy flux measured while the driving
term is applied is comparable to  
$E_T/t_d$, where $E_T$ is the steady-state value of the sum of
fluctuating magnetic and kinetic energies during the turbulent
driving phase, and $t_d$ is the dissipation time
measured after turbulent driving is discontinued.
Our obtained value $t_d \simeq 10t_0$, which does not depend
sensitively on most parameters, is somewhat longer than 
that estimated from three-dimensional periodic box simulations.
We believe that part of the discrepancy comes from
the generation of longer wavelength modes as the waves
travel to low-density regions near the cloud's surface.
However, one-dimensional simulations are also known to have 
lower dissipation rates than two- or three-dimensional ones
(Ostriker et al. 2001). Therefore, higher dimensional simulations 
that include density-stratification of the cloud will give 
the final answer in the future.

In our ideal MHD model there is no wave damping due to ion-neutral
friction, although there is wave damping due to a nonlinear cascade
and nonlinear steepening, as well as leakage of
wave energy into the external medium (see Paper I for a more
extensive discussion). We expect that ion-neutral friction
would be most effective at damping the shortest wavelength modes
in our simulation (Kulsrud \& Pearce 1969), 
and that the power spectra of $B_y$ 
and $v_y$ may become steeper at high wavenumber. However, our
current results show
that the wavelengths that are comparable to the 
initial cloud size are most effective at expanding the cloud,
so it is not clear exactly how effective ion-neutral friction
would be in removing turbulent cloud support. 
Furthermore, we are modeling the large scale evolution of molecular
clouds with mean column densities of up to a few times 10$^{21}
\, \cms$, or visual extinction $A_V \lesssim 3$. At these column 
densities, the clouds are expected to be significantly ionized by
background far-ultraviolet starlight so that ion-neutral friction 
will take a 
prohibitively long time (McKee 1989). Our results show
that the dense midplane of the cloud has transonic or subsonic
motions, and we expect these regions to decouple from the more
turbulent low density envelope and develop higher column density 
regions in which the far-ultraviolet background is shielded 
(McKee 1989).
In these regions, the ionization fraction is significantly lesser, 
and ambipolar diffusion can enhance the process of core formation.
Therefore, models of fragmentation of a thin-layer which include
the effects of magnetic fields and partial ionization due to 
cosmic rays (Basu \& Ciolek 2004) form a complementary approach to 
ours.

Our model has the advantage of being global by virtue of including
the effect of cloud stratification and a cloud boundary, rather
than being a local periodic box model. The 1.5-dimensional 
approximation has also allowed extremely high resolution which 
can more accurately track the propagation of MHD waves. 
Our localized turbulent driving source is an artificial
construct that mimics the various possible sources
of disturbances in a real cloud. However, a 
general result is that the wave energy gets quickly redistributed
throughout the cloud in such a way that most of the energy is on 
the largest scales. This result provides a way out of the 
``luminosity problem'' posed by Basu \& Murali (2001), since the
dissipation rate is then set by the crossing time of the cloud
scale rather than that of the internal driving scale.
Our modeled clouds have a steady-state size $\sim 1$ pc and
large-scale velocity dispersion $\sim 1 \kms$,  which makes them comparable 
to observed dark clouds which can form groupings known
as the giant molecular clouds (GMC's; see e.g., Blitz \& Williams 1999).
The turbulent dissipation typically occurs with an $e$-folding time of 
a few Myr and is completely dissipated by $\sim 10^7$ yr.
Even in the case that real molecular clouds and GMC's
start their life in a turbulent state and then evolve with {\em no}
further external turbulent input, our results show that the 
turbulent lifetime is less than but within reach of 
estimated GMC lifetimes, $\sim 10^7$ yr; see discussion in Larson 
(2003). This provides another important motivation for
future {\em global} three-dimensional models which can follow the
actual sources of internal turbulent driving.
As turbulence initially dissipates and a cloud contracts, 
we expect a transformation of some released gravitational potential energy
back into MHD wave energy. In addition, much of the internal driving
may be attributable to the outflows commonly launched
from the near environs of protostars (see e.g., Bachiller \& Tafalla 1999).

\acknowledgments

\clearpage

\centering
\begin{tabular}{ccccccccc} \hline
$\tilde{a}_d$ & $\beta_0$ & $\tilde{\nu}_0$ & $\lambda_0/H_0$ & $\langle z_F \rangle_t/H_0$ & $\langle \sigma_F^2 \rangle_t^{1/2}/c_{s0}
$ & $\langle z_H \rangle_t/H_0$ & $\langle \sigma_H^2 \rangle_t^{1/2}/c_{s0}$ & $t_d/t_0$ \\ \hline
10 & 1 & 1 & 1.41 & 4.00 & 2.03 & 0.760 & 1.09 & 9.0 \\
20 & 1 & 1 & 1.41 & 6.82 & 2.80 & 1.13  & 1.30 & 5.5 \\
30 & 1 & 1 & 1.41 & 10.8 & 3.42 & 1.78  & 1.53 & 8.5 \\
40 & 1 & 1 & 1.41 & 17.9 & 4.16 & 2.91  & 1.88 & 9.0 \\
50 & 1 & 1 & 1.41 & 17.4 & 4.34 & 3.34  & 2.02 & 15.0 \\ \hline

30 & 1 & 2    & 0.707 & 6.42 & 2.62 & 1.20 & 1.31 & 6.0 \\
50 & 1 & 2    & 0.707 & 8.54 & 3.01 & 1.78 & 1.44 & 5.5 \\
30 & 1 & 0.5  & 2.83  & 17.4 & 4.63 & 2.20 & 1.76 & 9.5 \\
20 & 1 & 0.5  & 2.83  & 6.60 & 2.73 & 1.09 & 1.32 & 6.5 \\ \hline

30 & 0.09 & 1 & 4.71  & 2.57 & 1.36 & 0.639 & 1.02 & 2.5 \\
30 & 0.16 & 1 & 3.54  & 2.88 & 1.54 & 0.750 & 1.02 & 3.0 \\
30 & 0.25 & 1 & 2.83  & 11.5 & 3.57 & 1.32  & 1.38 & 9.0 \\
20 & 0.25 & 1 & 2.83  & 5.33 & 2.47 & 0.86  & 1.11 & 8.0 \\
30 & 4    & 1 & 0.707 & 5.00 & 2.10 & 1.34  & 1.20 & 8.0 \\
50 & 4    & 1 & 0.707 & 6.06 & 2.19 & 1.94  & 1.23 & 8.0 \\ \hline

30 & 0.25 & 2   & 2.83  & 9.28 & 3.12 & 1.28  & 1.35 & 8.0 \\
30 & 4    & 2   & 0.354 & 3.22 & 1.60 & 0.837 & 1.06 & 8.0 \\
30 & 0.25 & 0.5 & 5.66  & 2.67 & 1.45 & 0.688 & 1.08 & 4.0 \\
30 & 4    & 0.5 & 1.41  & 8.52 & 2.76 & 2.68  & 1.59 & 8.5 \\ \hline

5000  & 1    & random & -- & 11.5 & 3.45 & 2.18  & 1.56 & 9.0 \\
5000  & 0.25 & random & -- & 11.8 & 3.88 & 1.56  & 1.44 & 6.5 \\
5000  & 4    & random & -- & 10.9 & 3.31 & 3.20  & 1.61 & 7.5 \\
10000 & 1    & random & -- & 22.3 & 4.90 & 4.16  & 1.91 & 7.5 \\
1000  & 1    & random & -- & 3.51 & 1.84 & 0.736 & 1.06 & 7.5 \\
100   & 1    & random & -- & 2.53 & 1.33 & 0.612 & 1.00 & 6.5 \\ \hline

\end{tabular}

\clearpage

\begin{figure}
\plotone{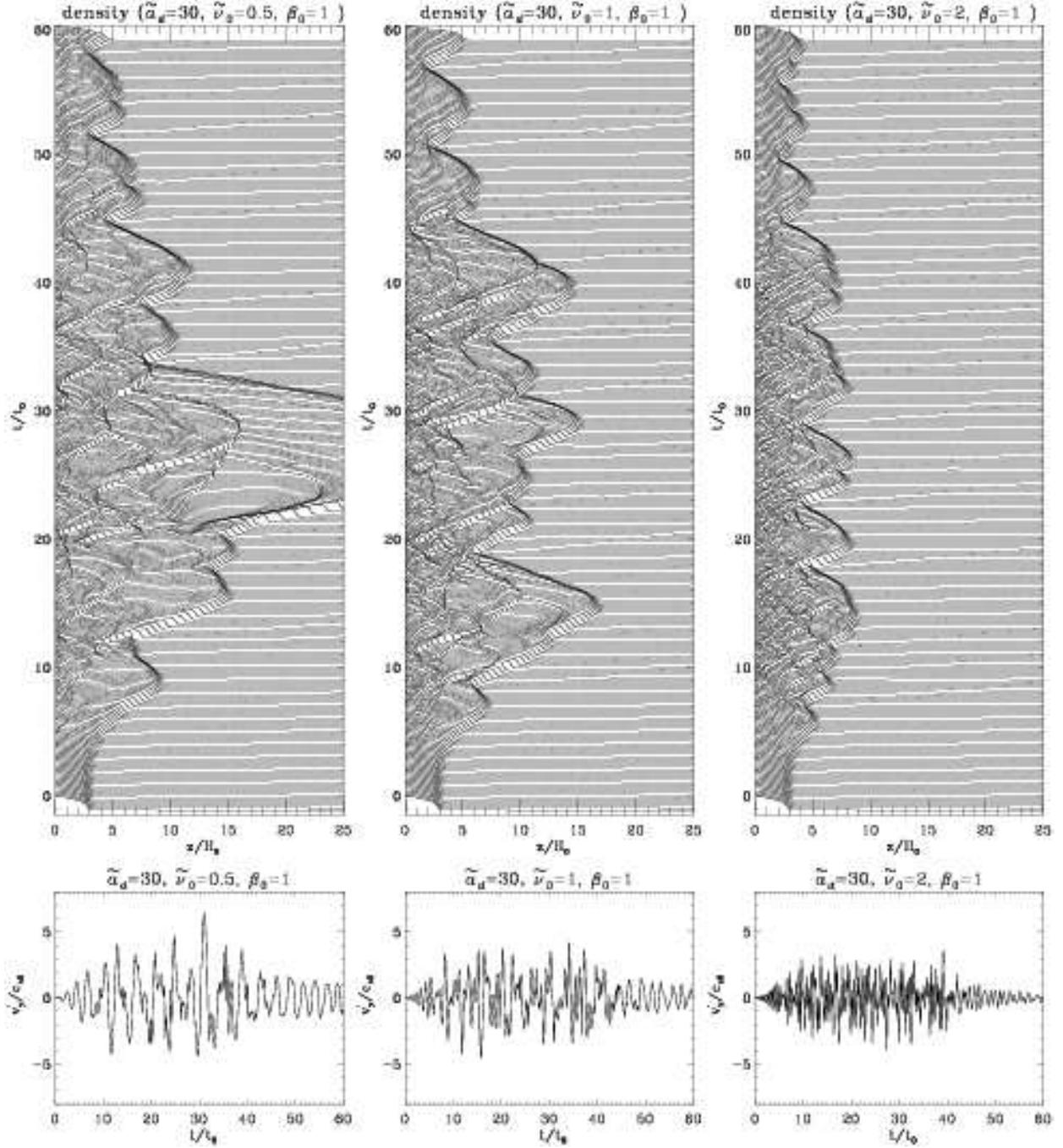}
\caption{
The time evolution for different frequencies
($\tilde{\nu}_0=0.5$, $\tilde{\nu}_0=1$, $\tilde{\nu}_0=2$)
with the same parameters of $\tilde{a}_d=30$ and $\beta_0=1$.
The upper panels show the time evolution of density.
The density profile at various times are stacked with time
increasing upward in uniform increments of $0.2t_0$.
The lower panels show the time evolution of $v'_y$
at $z=0$.
}
\end{figure}

\begin{figure}
\plotone{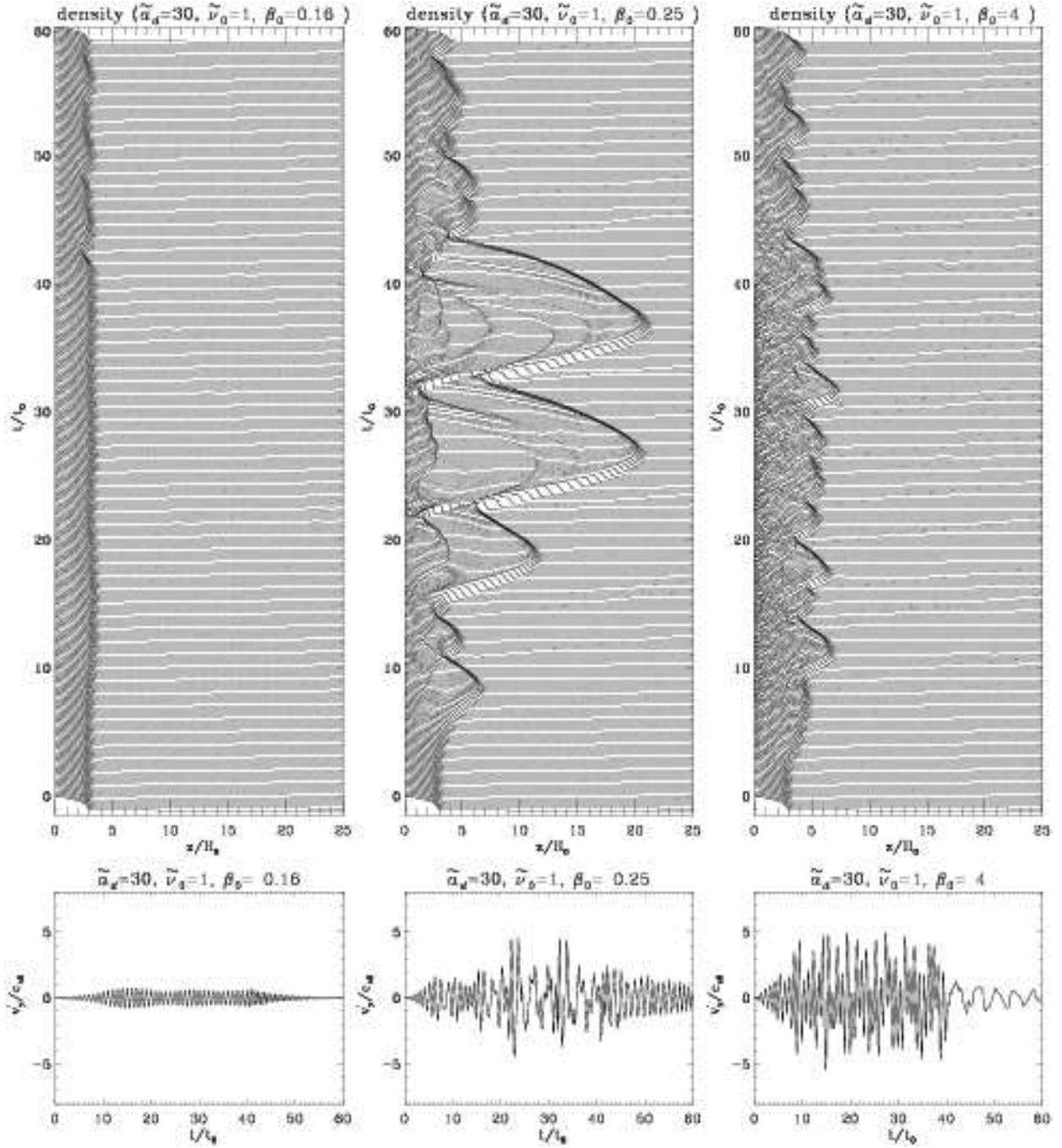}
\caption{
The same as Figure 1 but for different plasma beta
($\beta_0=0.16$, $\beta_0=0.25$, $\beta_0=4$)
with the same $\tilde{a}_d=30$ and $\tilde{\nu}_0=1$.
}
\end{figure}

\begin{figure}
\centering
\includegraphics[width=11cm]{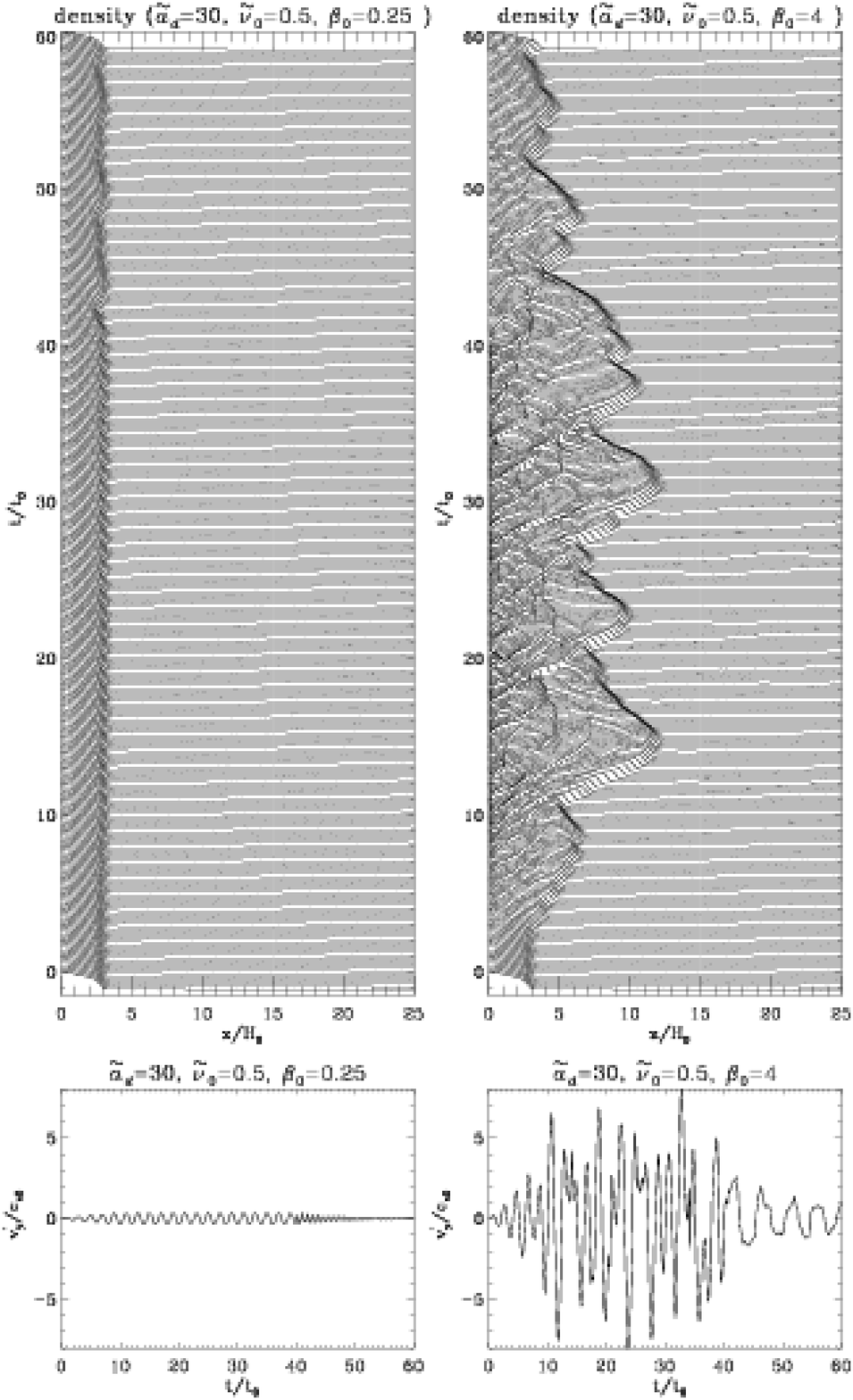}
\caption{
The same as Figure 1 but for different parameters.
The left panel shows the case of $\beta_0=0.25$,
$\tilde{\nu}_0=0.5$ and $\tilde{a}_d=30$. 
The right panel shows the $\beta_0=4$, $\tilde{\nu}_0=0.5$
and $\tilde{a}_d=30$.
}
\end{figure}

\begin{figure}
\centering
\includegraphics[width=11cm]{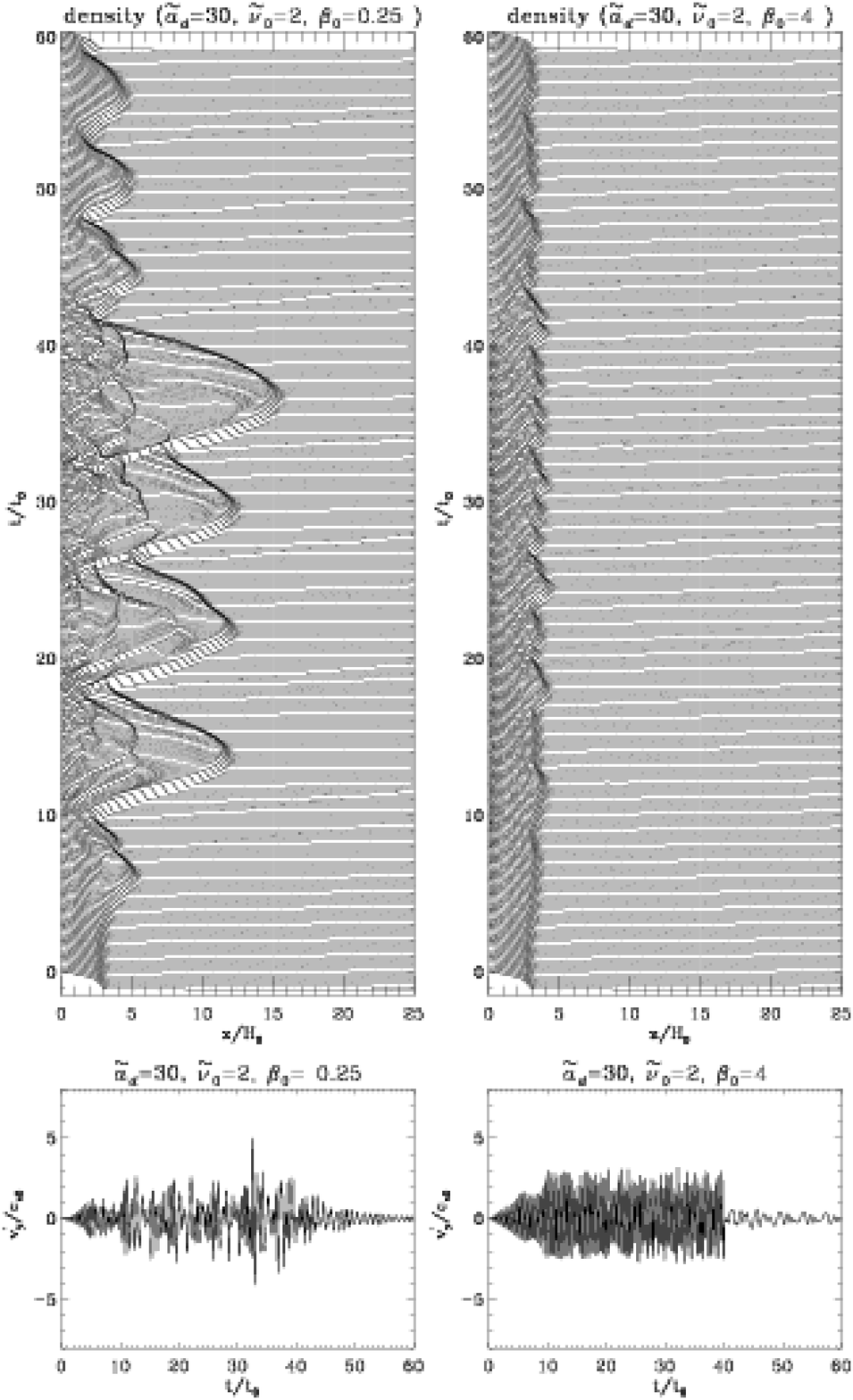}
\caption{
The same as Figure 1 but for different parameters.
The left panel shows the case of $\beta_0=0.25$,
$\tilde{\nu}_0=4$ and $\tilde{a}_d=30$. 
The right panel shows the $\beta_0=4$, $\tilde{\nu}_0=2$
and $\tilde{a}_d=30$.
}
\end{figure}

\begin{figure}
\plotone{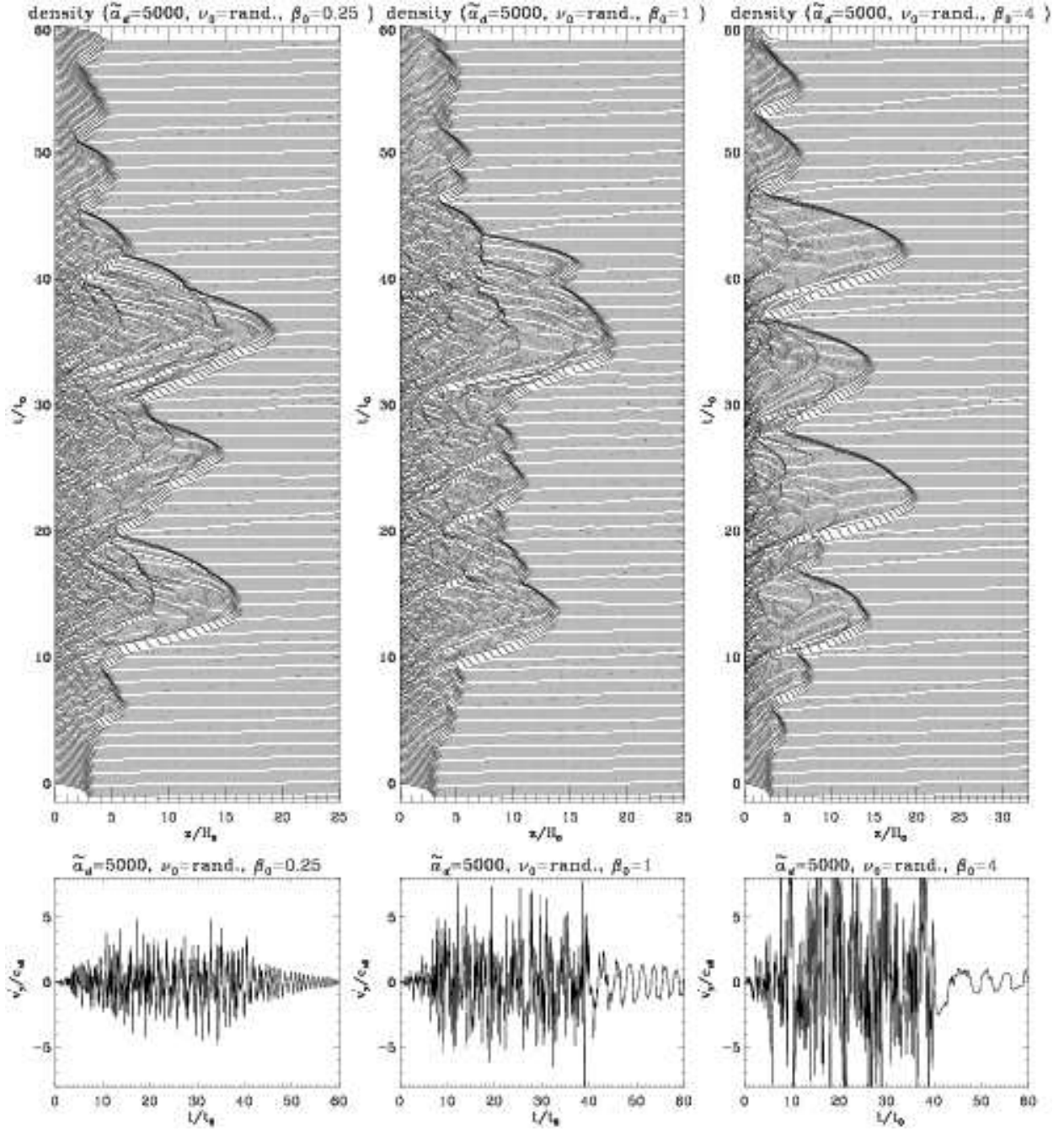}
\caption{
The time evolution for random driving case with $\tilde{a}_d=5000$.
The left panel shows the case of $\beta_0=0.25$,
the middle panel shows $\beta_0=2$, and the right panel shows 
$\beta_0=4$.
}
\end{figure}

\begin{figure}
\plotone{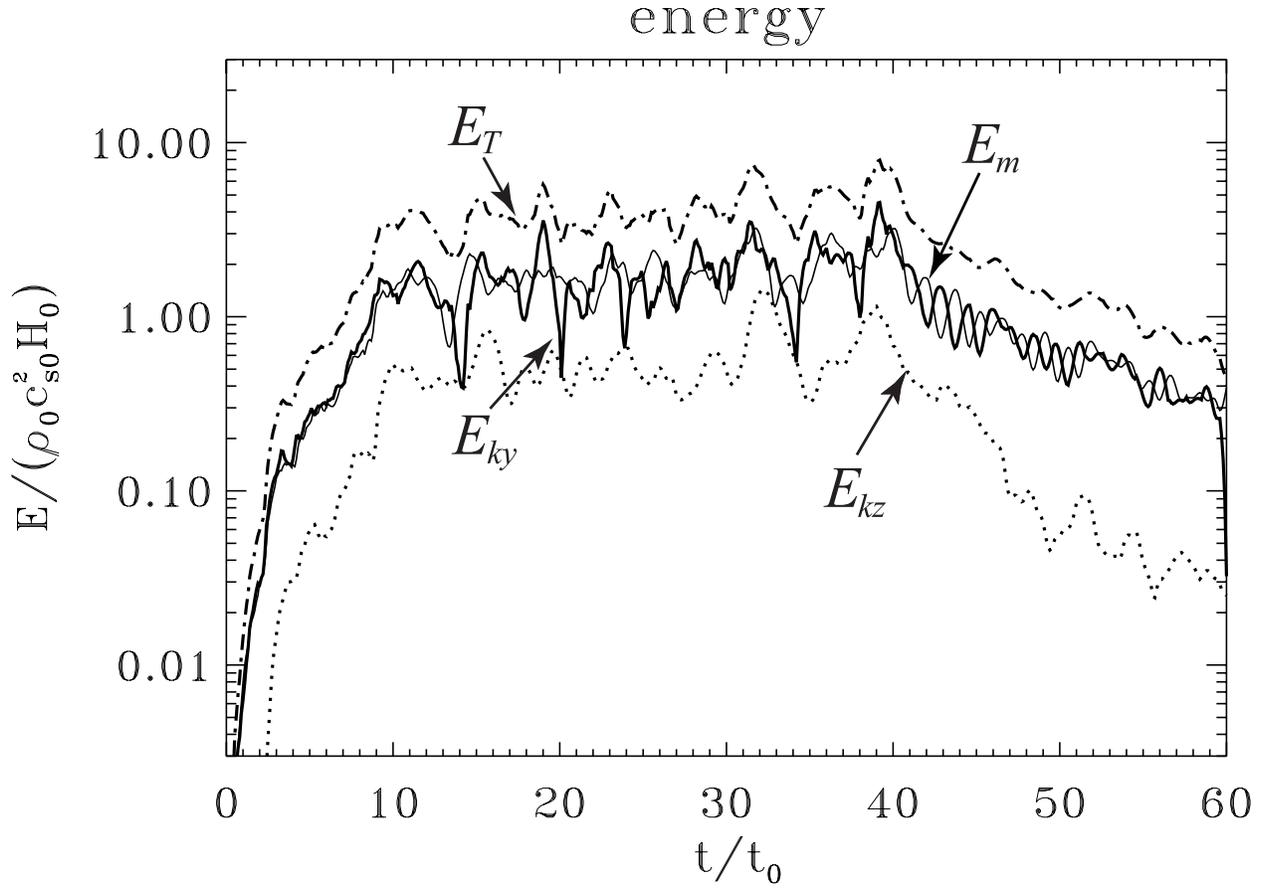}
\caption{
Time evolution of various energies in the cloud for random
driving force with $\tilde{a}_d=5000$ and $\beta_0=1$.
The thick solid line shows $E_{ky}$, the dotted line shows $E_{kz}$,
the thin solid line shows $E_{m}$, and dash-dotted line shows $E_{T}
= E_{ky} + E_{kz} + E_{m}$. 
The values of each energy are smoothed out over a time $t_0$ in order 
to remove small oscillations originating from the driving force.
}
\end{figure}

\begin{figure}
\plotone{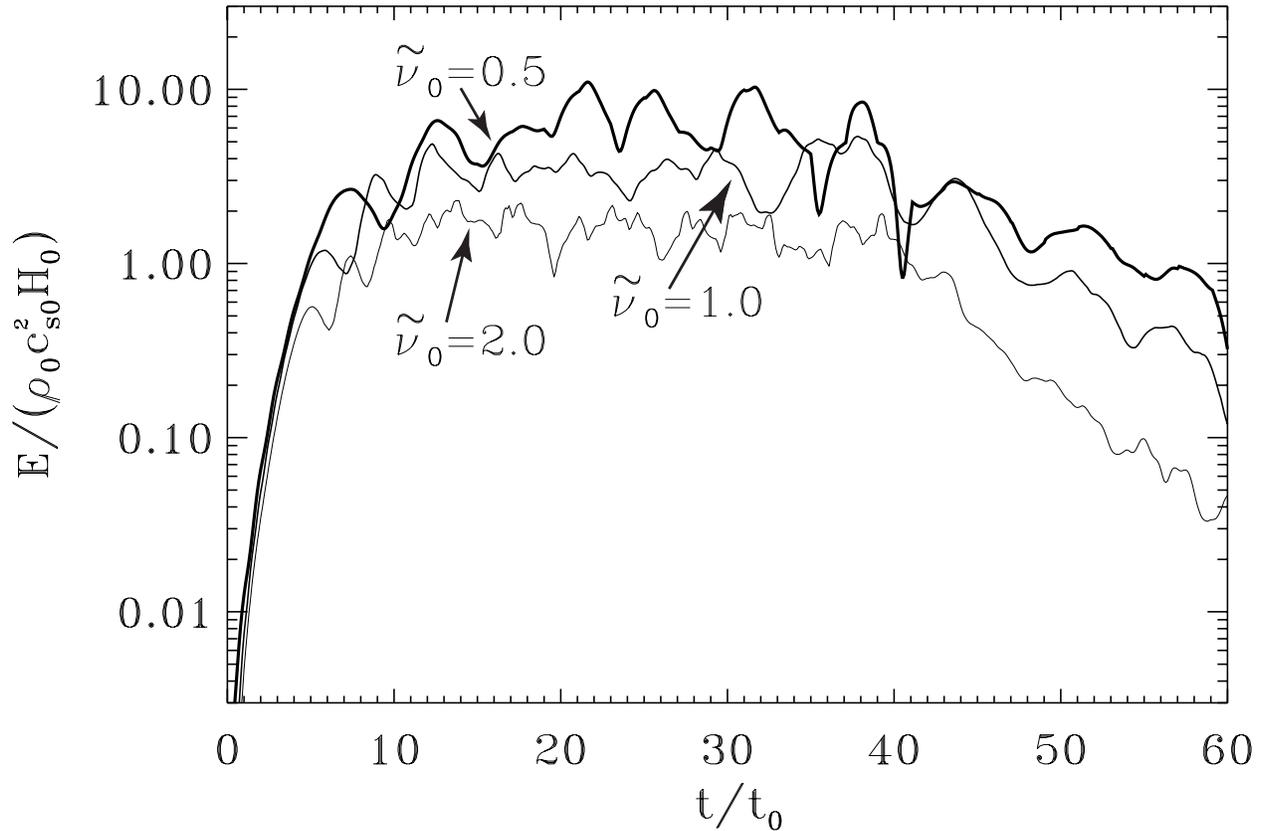}
\caption{
Time evolution of $E_{T}
= E_{ky} + E_{kz} + E_{m}$ in the cloud for different
frequencies ($\tilde{\nu}_0=0.5$, $\tilde{\nu}_0=1$, $\tilde{\nu}_0=2$)
with the same parameters of $\tilde{a}_d=30$ and $\beta_0=1$.
The thick solid line shows the case of $\tilde{\nu}_0=0.5$,
the solid line shows $\tilde{\nu}_0=1$, and the thin solid line 
shows $\tilde{\nu}_0=2$. The values of each energy are smoothed out 
over every one cycle of the driving force in order to remove 
small oscillations originating from the driving force.
}
\end{figure}

\begin{figure}
\plotone{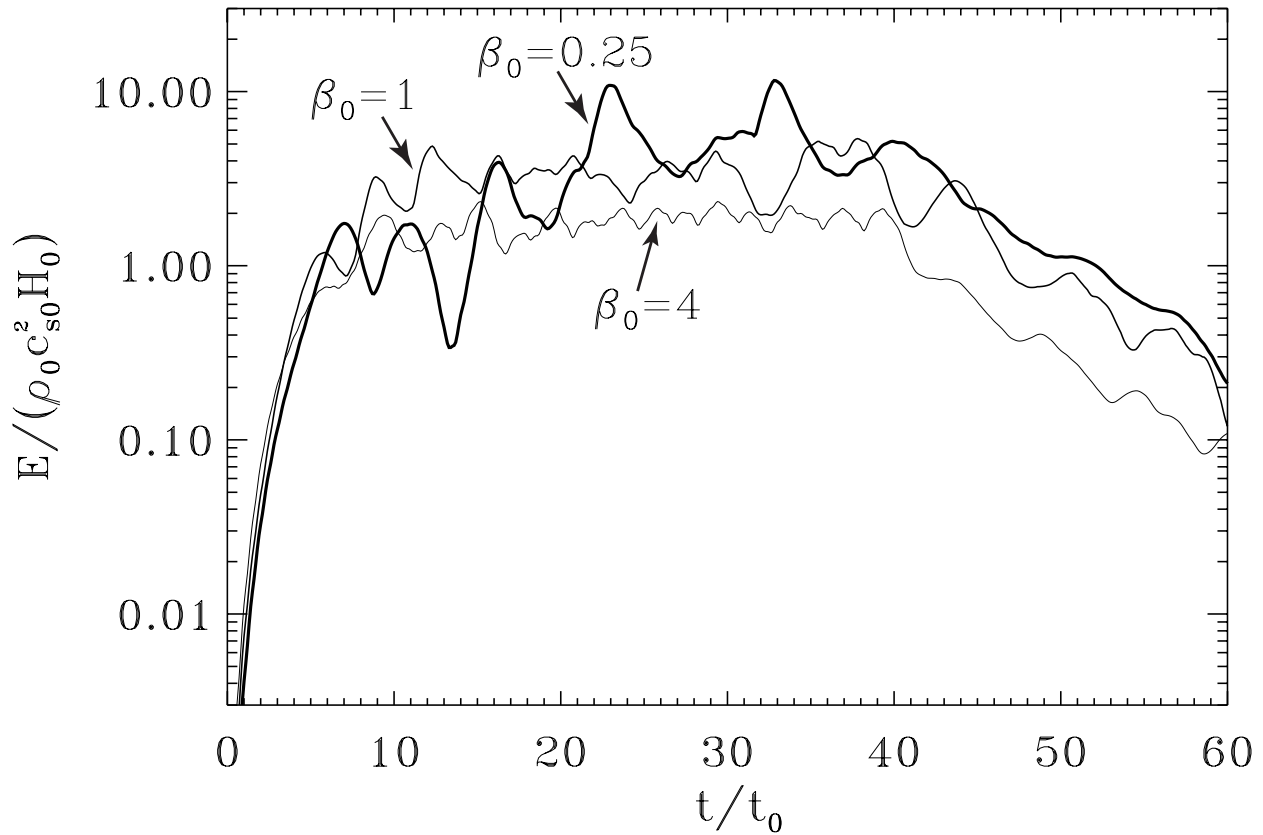}
\caption{
Time evolution of total energies in the cloud for different
plasma beta ($\beta_0=0.25$, $\beta_0=1$, $\beta_0=4$)
with the same parameters of $\tilde{a}_d=30$ and $\tilde{\nu}_0=1$.
The thick solid line shows the case of $\beta_0=0.25$,
the solid line shows $\beta_0=1$, and the thin solid line 
shows $\beta_0=4$. The values of each energy are smoothed out 
over every one cycle of the driving force in order to remove 
small oscillations originating from the driving force.
}
\end{figure}

\begin{figure}
\plotone{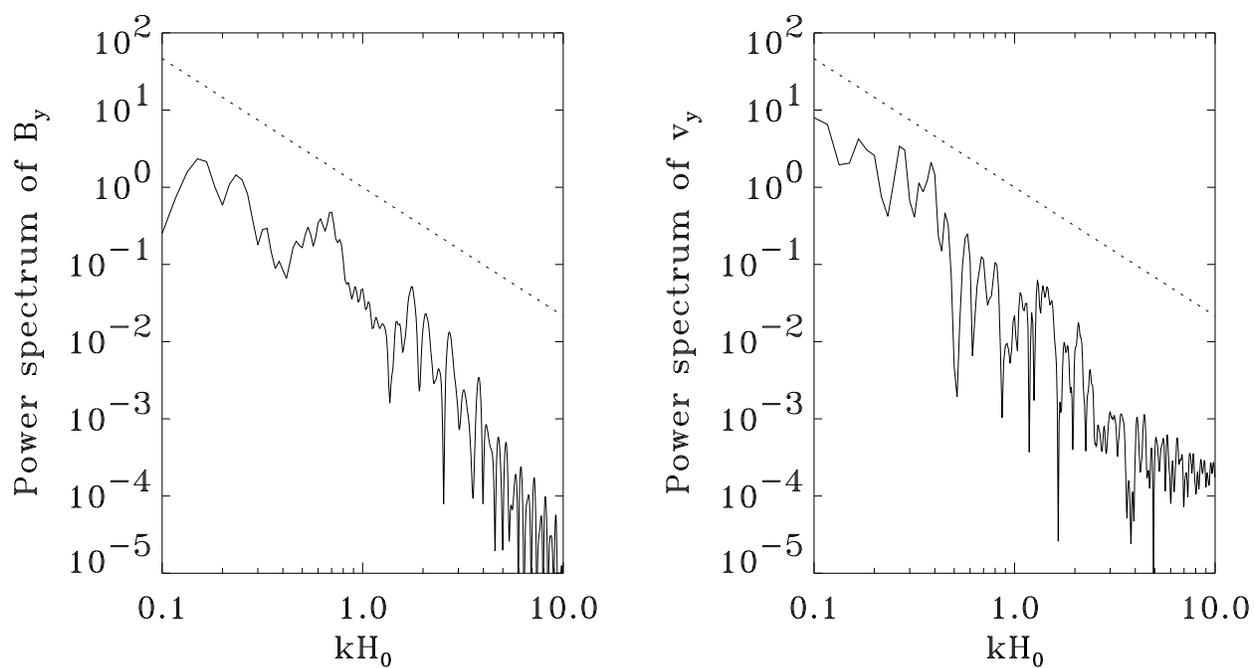}
\caption{
Spatial power spectra of $B_y$ and $v_y$ for the case of
the sinusoidal driving force with parameters $\tilde{a}_d=30$, 
$\tilde{\nu}_0=1$ and $\beta_0=1$. The spectra were taken from 
a snapshot at $t=30t_0$. The dotted line is proportional 
to $k^{-5/3}$ and is shown for comparison.
}
\end{figure}

\begin{figure}
\plotone{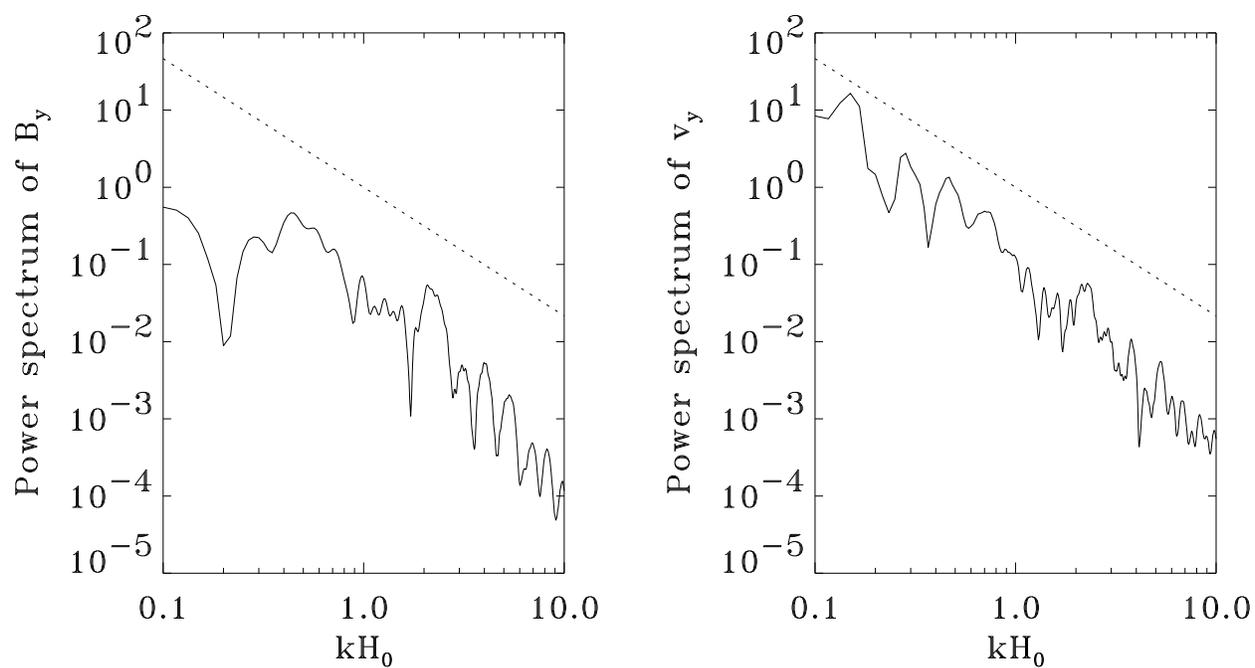}
\caption{
Spatial power spectra of $B_y$ and $v_y$ for the case of
the random driving force with parameters 
$\tilde{a}_d=5000$ and $\beta_0=1$. 
The spectra were taken from the snapshot at $t=30t_0$. 
The dotted line is proportional to 
$k^{-5/3}$ and is shown for comparison.
}
\end{figure}

\begin{figure}
\plotone{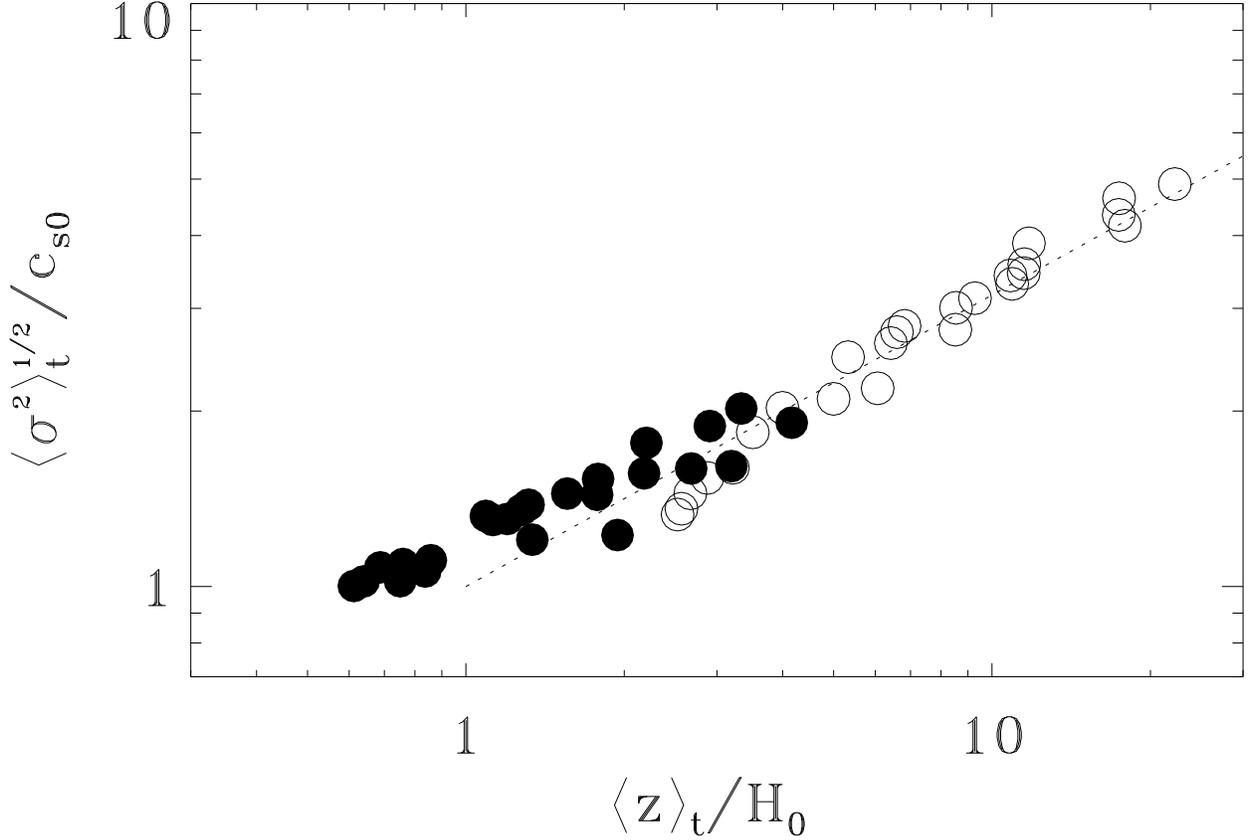}
\caption{
Time averaged velocity dispersions of different Lagrangian fluid 
elements for different parameters as a function of time averaged 
positions. The open circles correspond to Lagrangian fluid elements
whose initial positions are located at $z=2.51 H_0$,
which is close to the edge of the cold cloud.
The filled circles correspond to Lagrangian fluid elements
whose initial positions are located at $z=0.61 H_0$,
which is approximately the half-mass position of the cold cloud.
The dotted line shows 
$\langle \sigma^2 \rangle_t^{1/2} \propto \langle z \rangle_t^{0.5}$.
Each circle can be associated with a particular model in our study by
comparison with the numbers in Table 1.
}
\end{figure}

\begin{figure}
\plotone{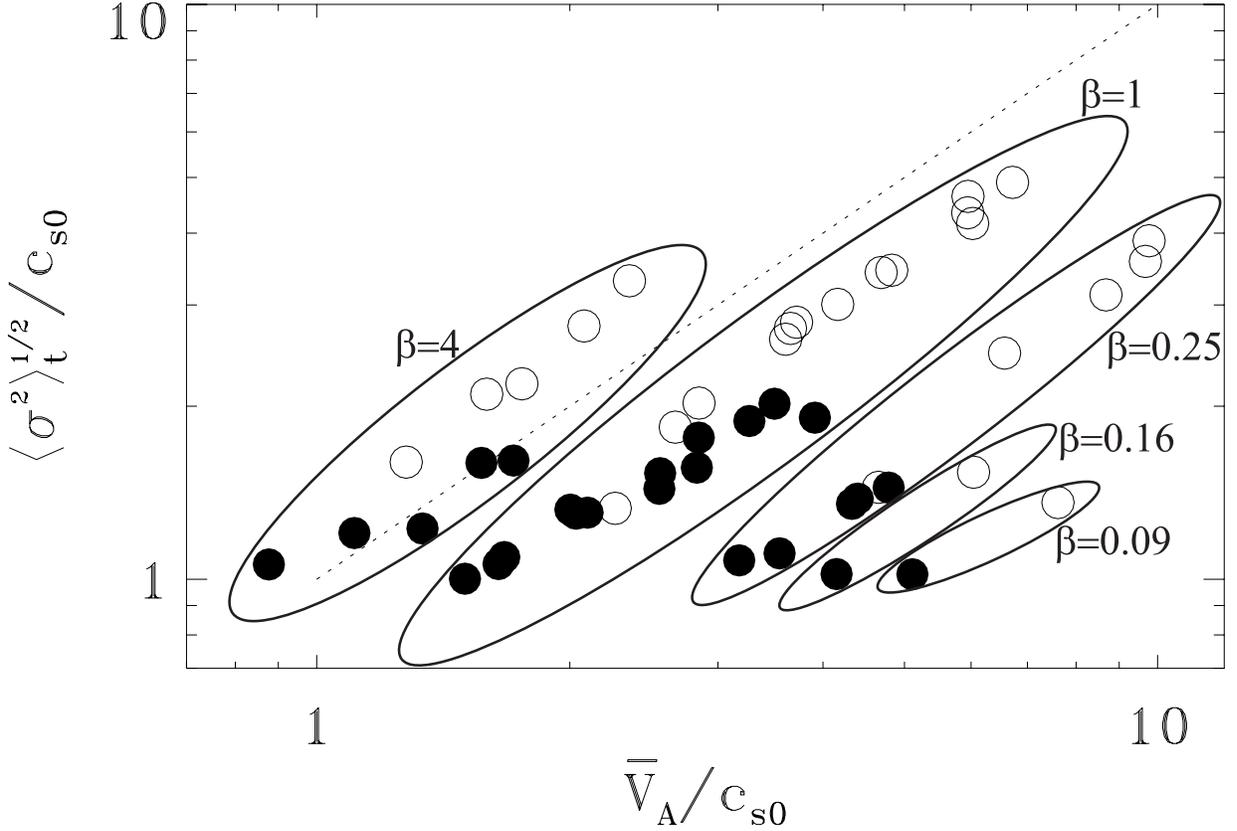}
\caption{
Time averaged velocity dispersions as a function of 
the mean Alfv\'en velocity of the cloud.
The results from the same plasma beta are encircled.
The open circles correspond to Lagrangian fluid elements
whose initial positions are located at $z=2.51 H_0$.
The filled circles correspond to Lagrangian fluid elements
whose initial positions are located at $z=0.61 H_0$.
The dotted line shows  
$\langle \sigma^2 \rangle_t^{1/2} \propto \bar{V}_A$.
}
\end{figure}

\end{document}